\documentclass[10pt]{article}
\usepackage[]{emulateapj}
\usepackage[]{psfig}

\begin{document}
\submitted{}

\title{Strong, Variable Circular Polarization in PKS~1519-273}

\author{J-P Macquart\altaffilmark{1}, 
Lucyna Kedziora-Chudczer\altaffilmark{1,2}, David P. Rayner\altaffilmark{2,3}, 
David L. Jauncey\altaffilmark{2}}
\affil{$^1$Special Research Centre for Theoretical Astrophysics, School of 
Physics, University of Sydney, New South Wales 2006, Australia.}
\affil{$^2$Australia Telescope National Facility, CSIRO, Epping, New 
South Wales 2121, Australia.}
\affil{$^3$School of Mathematics and Physics, University of Tasmania, 
Hobart, Tasmania 7001, Australia.}


\begin{abstract}
We report strong variability in the circular and linear 
polarization of the intraday variable source PKS~1519-273.
The circular polarization varies on 
a timescale of hours to days at frequencies between 1.4 and 8.6~GHz, and 
is strongly correlated with variations in the total intensity at 4.8 
and 8.6~GHz.  We argue that the variability is due to 
interstellar scintillation of a highly compact 
($15-35 \mu$as) component of the 
source with $-3.8\pm 0.4$\% circular polarization at 4.8~GHz.
We find that no simple model for the circular polarization 
can account for both the high magnitude and the frequency 
dependence in PKS~1519-273 at centimeter wavelengths.  
\end{abstract}


\section{Introduction}

Circular polarization (CP) in extragalactic sources is very small, typically 
0.05\% to 0.1\% of the total source flux density (e.g. 
Roberts et al. 1975, Seaquist et al. 1974, Weiler and de Pater 
1983)\markcite{Roberts75,Seaquist74,Weiler83} and sometimes variable 
(Komesaroff et al., 1984). Recent measurements of 
CP in extragalactic sources have
rekindled debate as to its characteristics and origin.  
Wardle et al. (1998)\markcite{Wardle98} detected CP 
in 3C~279 and attributed it to the presence of a relativistic pair plasma. 
New evidence has emerged that Sgr A$^*$, the AGN-like object at the 
core of our own Galaxy, is also weakly circularly polarized (Bower et al. 
1999, Sault \& Macquart 1999)\markcite{Bower99,Sault99}.

We present Australia Telescope Compact Array (ATCA) measurements of the 
timescale and magnitude of the variability of the CP 
in the extragalactic, intraday variable (IDV) BL Lac object,   
PKS~1519-273 (White et al. 1988)\markcite{White88}.  
PKS~1519-273, at Galactic co-ordinates $l=339.5^{\circ}, \, b=24.5^{\circ}$,
is identified with a $m_V=18.5$ star-like object with a 
featureless optical spectrum. The lower limit on its redshift is $z=0.2$ 
(Veron-Cetty \& Veron 1993)\markcite{Veron93}.  
PKS~1519-273 is a compact high brightness temperature 
radio source (Linfield et al. 1989)\markcite{Linfield89}.  The ATCA IDV
Survey data shows strong IDV (Kedziora-Chudczer 
1998)\markcite{KedzioraPhD} and IDV of the total and polarized 
flux densities at GHz frequencies has been 
found during each of the 5 epochs of ATCA observations over the 
past 7 years.

PKS~1519-273 has not been seen to exhibit IDV at 
either optical (Heidt \& Wagner 1996)\markcite{Heidt96} or mm 
wavelengths (Steppe et al. 1988, 
Steppe et al. 1995)\markcite{Steppe88,Steppe95}.  However, it does have a 
high degree ($5-12$\%) of variable optical linear polarization (Impey and Tapia 
1988)\markcite{Impey96}.  PKS~1519-273 is a weak, soft X-ray source 
with a flux density at 1~keV of 0.39~$\mu$Jy 
(Urry et al. 1996)\markcite{Urry96}. Its $\gamma $-ray 
energy output is less than 
$0.7\times 10^{-7}$photons~cm$^{-2}$sec$^{-1}$ for energies 
$E>100$~MeV (Fichtel et al. 1994)\markcite{Fichtel94}.

\section{Observations and Results}
We base our present report on PKS~1519-273 on the data obtained with ATCA 
over 5 days starting on 1998 September 9.  Data were collected simultaneously for two 
frequencies centered on either 1.384 and 2.496 GHz, or 4.800 and 
8.640 GHz each with a 128 MHz bandwidth. 
To ensure high quality amplitude and phase calibration
we frequently observed both the standard primary flux density calibrator, 
PKS~1934-638, and a secondary calibrator, PKS~1514-241. 
The primary calibrator was used to determine accurately the flux 
density scale and the instrumental 
polarization leakages (e.g. Sault, Killeen \& Kesteven 1991)\markcite{Sault91}.  The 
total and polarized flux density lightcurves of 
PKS~1519-273 are presented in fig. 1. The circularly polarized emission, is unresolved on all ATCA baselines and is strongly variable at 4.8 and 
8.6~GHz.  Comparison of the 2.5, 4.8 and 8.6~GHz Stokes $V$ measurements for 
PKS~1519-273 and the strong calibrator source PKS~1514-241, extensive 
testing and consistency checks demonstrate that the observed CP and its 
variations are not instrumental effects. 

The most striking features of the 4.8 and 8.6~GHz light curves in 
fig. 1 are the exceptionally high level of CP 
and the large amplitude variability in all four Stokes 
parameters.  The fractional variability of both the circularly and 
linearly polarized flux density exceeds that of the total flux 
density.  The high degree of correlation between the fluctuations in $I$ and 
$V$ (see figs. 1 \& 2) suggests that the mechanism of variability of the 
CP is strongly related to that in $I$.  Comparison of the 
fluctuations in $V$ with those in $I$ implies that, although the 
overall CP is only $\sim 1$\%, the
CP of the variable component, $\Delta V/\Delta I$, is 
$-2.4 \pm 1.3$\%, $-3.8 \pm 0.4$\% and $-2.6 \pm 0.5$\% at 2.4, 4.8 and 
8.6~GHz respectively (see figs. 2 \& 3(c)).  The CP is 
weaker, $< 1.3$\% at 1.4~GHz and its variability is less well-established.

\placefigure{fig:HIGH.PS}
\begin{figure*}
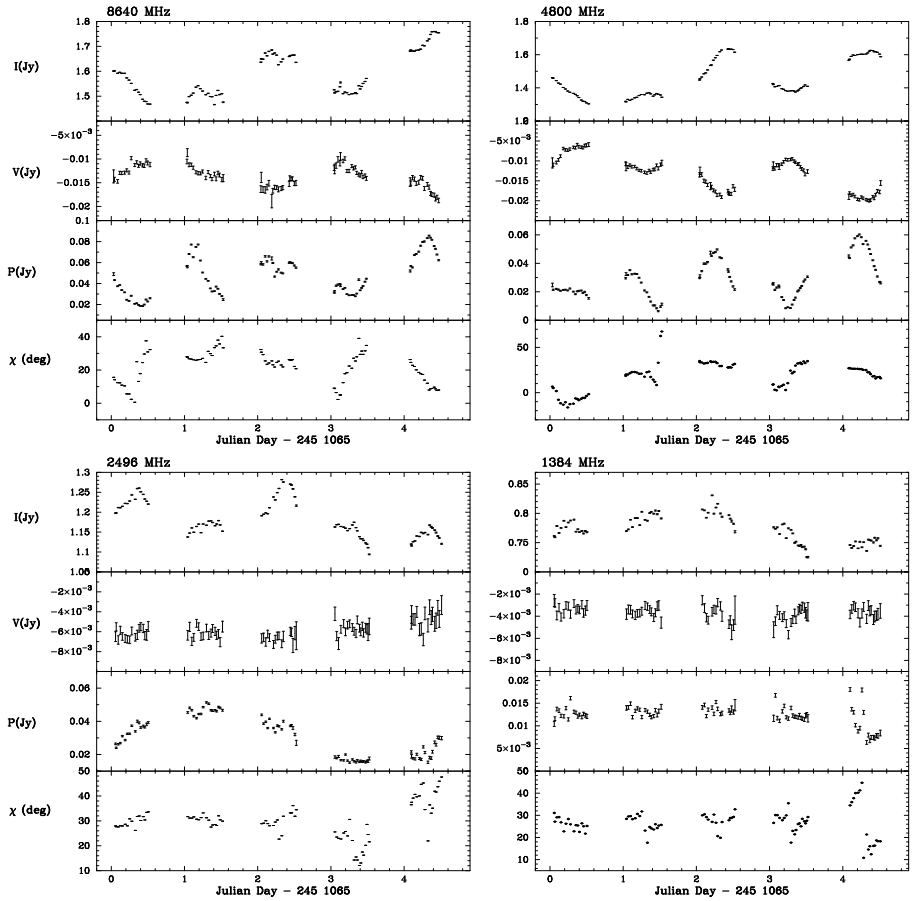

    \begin{center}
\begin{tabular}{c}
    \psfig{file=HIGH.PS,angle=270,width=120mm} \\
    \psfig{file=LOW.PS,angle=270,width=120mm}
\end{tabular}
\end{center}
\caption{Variability of PKS~1519-273 in the total intensity ($I$), 
CP ($V$), and magnitude of the linear polarization ($P$),
for the four bands of the ATCA over 5 days.  Each point 
represents a 10~min average, and is plotted with $1\,\sigma$ error bars. 
Fluctuations in $I$ and $V$ are strongly correlated ($r=-0.90$ at 
4.8~GHz and $r=-0.80$ at 8.6~GHz).}
\end{figure*}

\placefigure{fig:v_vs_i_2496.eps}
\begin{figure*}
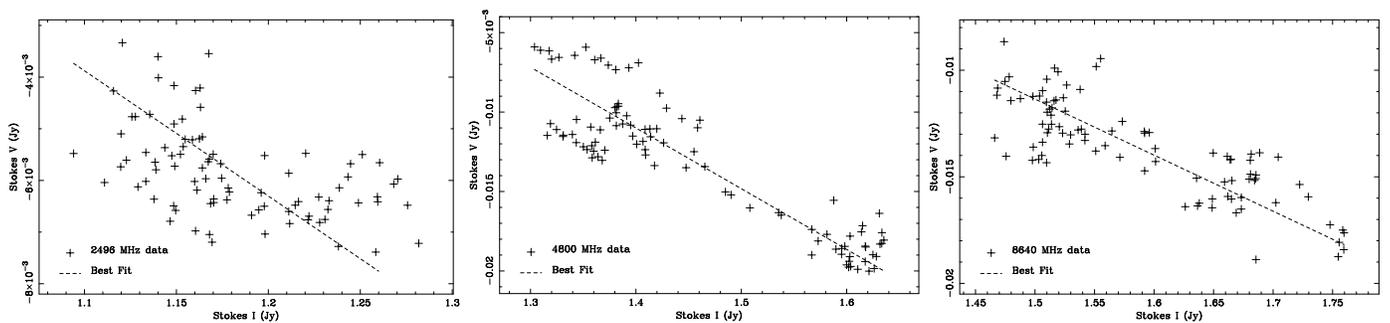

\begin{tabular}{c}
    \psfig{file=v_vs_i_2496.eps,angle=270,width=60mm} 
    \psfig{file=v_vs_i_4800.eps,angle=270,width=60mm} 
    \psfig{file=v_vs_i_8640.eps,angle=270,width=60mm}
\end{tabular}
\caption{Plot 
of the correlation between $V$ and $I$.  The best fit lines give 
a fractional CP of $-2.4\pm 1.3$\%, $-3.8 \pm 0.4$\% 
and $-2.6 \pm 0.5$\% respectively at 2.5, 4.8 and 8.6~GHz for the 
variable component of the source.   The fluctuations in $I$ and $V$ do 
not appear to be correlated at 1.4~GHz, indicating that we detect 
{\it no} CP associated with the variable component 
at this frequency.  We place an upper limit on the CP of 
this component of 1.3\%.}
\end{figure*}

\section{Discussion}
\subsection{Scintillation}

We attribute the short-timescale variability of this source to 
Interstellar Scintillation (ISS) in our Galaxy.  ISS has already been 
invoked to explain radio source IDV (Heeschen \&
Rickett 1997)\markcite{Heeschen97}, 
including the rapid variability of PKS~0405-385 (Kedziora-Chudczer et 
al. 1997)\markcite{Ked97}.  If intrinsic to the source, the total intensity 
variations observed at 4.8~GHz imply a brightness temperature of $T_B \gtrsim 3 
\times 10^{17}$~K for $z \gtrsim 0.2$, based simply on light travel 
times.  However, assuming that the 
variations are intrinsic implies a source size that is necessarily 
sufficiently small to exhibit variability due to 
ISS (Rickett et al. 1995)\markcite{Rickett95}.  This suggests that an 
explanation based on ISS should be sought first.

The increase in modulation index (the rms normalized by the mean intensity), shown in fig. 3, and
the short variability timescale from 8.6 to 4.8~GHz, shown in fig. 1, 
are consistent with 
scintillation in the regime of weak scattering 
(e.g., Narayan 1992)\markcite{Narayan92}, while the decrease in 
modulation indices and the increasingly longer 
variability timescales from 4.8 to 1.4~GHz 
are characteristic of refractive scintillation.

Assuming that the density 
inhomogeneities in the ISM are located on a thin screen, the refractive scintillation at 1.4~GHz may be used to place a 
constraint upon the distance to the scattering screen.  The physical 
extent of the scattering disk at 1.4~GHz is the product of the long-period,
refractive variability 
timescale, $t_{1.4}$, no 
less than 4 days (see fig. 1), and 
the scintillation speed, $v$, of order 50~km/s (see Rickett et al. 1995).  VSOP
observations at 1.7~GHz\footnote{See 
http://www.vsop.isas.ac.jp/general/pr/1519-273.gif} indicate that the 
source is unresolved, so we assume an
angular size of no more than 0.3~mas.  This implies an observer-screen distance of 
$D \geq 390 (v/50~{\rm km/s}) (t_{1.4}/4~{\rm days})$~pc, or 390~pc in the
present case.

Having obtained a lower limit to the distance to the 
scattering screen, we may constrain the angular diameter of the source 
from the scintillation parameters in the weak scattering regime where 
the scattering is quite sensitive to source size effects.  The 
scintillation timescale of $\sim 12$~hours at 4.8~GHz can be explained 
either in terms a scattering screen at large distance ($> 15$~kpc), or 
by a partially resolved source.  For weak 
scattering, the source is resolved if the angular diameter of the 
source, $\theta_S$, exceeds the angular diameter of 
the first Fresnel zone $\theta_F = (k D)^{-1/2}$, where $k$ is the 
wavenumber.  The 
scintillation timescale is then $t_{4.8} \approx \theta_S D/v$ for $\theta_S >
\theta_F$ (Narayan 1992).
A screen in our own Galaxy implies $D \ll 15$~kpc, so the source must be partially resolved.  
Assuming the asymptotic results of weak scattering to be valid 
between the weak and strong scattering regimes, the scintillation timescale  
then yields an {\it estimate} of the intrinsic angular source size of 
$$
\theta_S \approx 14.4 \left(\frac{t_{4.8}}{12~{\rm hours}} \right) 
\left( \frac{v}{50~{\rm km/s} } \right) 
\left( \frac{D}{1~{\rm kpc}}\right)^{-1}~\mu{\rm as}. \eqno(1)
$$

For a scintillating source of flux density $I_0$, the root-mean-square 
fluctuation is $I_{\rm rms} = I_0 m(\theta_S)$, where 
$m(\theta_S)=(kD \theta_S^2)^{7/12}$ is the modulation index expected for 
a source of size $\theta_S > \theta_F$ (e.g. Narayan 1992)\markcite{Narayan92}, 
and we have assumed that 4.8~GHz is near the 
transition frequency between weak and strong scattering.
Given $I_{\rm rms} = 0.11$~Jy and with the derived angular size of the 
scintillating component of the source, we estimate $I_0$ and derive 
its brightness temperature:
{\small
$$
T_b \approx 2.0 \times 10^{14} \left( \frac{D}{1~{\rm kpc}} 
\right)^{17/12} 
\left( \frac{t_{4.8}}{12~{\rm hours}} \right)^{-5/6} 
\left( \frac{v}{50~{\rm km/s}} \right)^{-5/6}~{\rm K}. \eqno(2)
$$}
Using the limit of the distance to the scattering screen, the maximum 
possible angular size of the source for $t_{4.8}=12$~hours and 
$v=50$~km/s is $37\,\mu$as, the minimum brightness 
temperature is $T_b=5 \times 10^{13}$~K, consistent with incoherent synchrotron 
emission subject to relativistic beaming with a Doppler 
boosting factor $\delta \gtrsim 200 (1+z)$ 
(Readhead 1994)\markcite{Readhead94}.

However, if the CP observed at 
4.8~GHz is entirely due to the variable component, we may further 
constrain $T_b$.  From fig. 2 we have 
$I_0 = 0.35 \pm 0.04$~Jy, implying $m(\theta_S) \approx 0.32$ 
(consistent with the modulation index observed in $V$: $0.32$) and hence an 
angular size of $9.8 (D/1\, {\rm kpc})^{-1/2} \,\mu$as.  Comparing with 
equation (1), we have $v = 34 \,(D/1\,{\rm kpc})^{1/2}$~km/s, implying 
a brightness temperature of 
$3 \times 10^{14} (D/1\,{\rm kpc}) (t_{4.8}/12\,{\rm hours})^{-5/6}$~K.  
A scintillation speed exceeding $50$~km/s therefore implies 
a brightness temperature $T_b \gtrsim 6 \times 10^{14}$~K.

\subsection{Circular Polarization}
We consider the origin of the CP in terms of intrinsic synchrotron 
(Legg \& Westfold 1968)\markcite{Legg68}, partial conversion of the linear 
polarization into CP due to ellipticity of the 
natural wave modes of the cold background plasma (Pacholczyk 
1973)\markcite{Pac73} or 
of the relativistic electron gas itself (e.g. Sazonov 1969, 
Jones \& O'Dell 1977a,b)\markcite{Saz69,Jones77a,Jones77b}.

If the CP in 
the scintillating component is due to 
synchrotron emission then, following Legg \& Westfold~(1968)\markcite{Leg68}, 
the Lorentz factor of the particles responsible for a CP 
of $m_c$ in a {\it uniform} magnetic field 
is $\gamma \approx \cot \theta/m_c$, where 
$\theta$ is the angle between the magnetic field
and the line of sight.  For $30^\circ < \theta < 60^\circ$ this 
implies Lorentz factors in the range $\approx 15-45$ to
explain the CP at 4.8~GHz. Indeed, the observed (low) 
level of linear polarization suggests a
non-uniform magnetic field, indicating that 
even higher Lorentz factors are required.  The 
maximum brightness temperature of such emission is $T_B \leq 
5.9 \times 10^9 \gamma$~K$=2.7 \times 10^{11}$~K in the rest frame. Bulk motion with a 
Doppler boosting factor $\delta \gtrsim 200$ would account for the difference 
between the rest-frame and scintillation-derived brightness 
temperatures.  Such a Doppler factor, although very high, cannot be 
entirely ruled out.  However the observed frequency dependence of the degree of 
CP is far from 
the $\nu^{-1/2}$ expected from synchrotron theory (see fig 3c); the 
CP {\it decreases} sharply between 8.6 and 1.4~GHz.
It is therefore unlikely that the CP is due to synchrotron emission.

Alternatively, the CP could be due to propagation 
through a relativistic pair plasma, such as may be present within the 
source itself.  The 
birefringence induced in a medium by the presence of a pair plasma 
may convert linear polarization to CP as 
follows:
$$
V(\nu) = U_{0}(\nu) \sin ( c^3 {\rm RRM}/\nu^3), \eqno(3)
$$
where the relativistic rotation measure, RRM, depends upon the 
density of relativistic particles, the path length, the magnetic 
field, and the minimum Lorentz factor of the pairs (e.g. Kennett \& Melrose 
1998)\markcite{Ken98}.
This effect operates only when the direction of the incident linear 
polarization is at an oblique angle to the projection of the magnetic 
field on the plane orthogonal to the ray direction.  The axes used to 
define the Stokes parameters may be chosen such that synchrotron 
emission has $Q \neq 0, U=0$.  With this choice, the effect occurs 
only if the incident radiation has $U_0 \neq 0$, requiring either 
Faraday rotation or that it 
originate from a region of the source where the magnetic field is in a 
different direction to that where the polarization 
conversion takes place.  

A characteristic of this model is a strong frequency dependence on the 
sign of $V$.  If ${\rm RRM}$ is high enough to produce the CP 
observed at high frequency, this model predicts rapid changes 
in $V$ at low frequency: in particular, equation (3) yields the lower limit 
$|{\rm RRM}| \geq 6.2 \times 10^2 \, (1+z)^3
/f_U(8.6\,{\rm GHz})$~rad/m$^3$  
to explain the $-2.6$\% CP at 8.6~GHz, where we write $f_U(8.6\,{\rm GHz}) =
|(U_0(8.6\,{\rm GHz})/I(8.6\,{\rm GHz})|$.  
For this lower limit, $\lambda^3 {\rm RRM}$ will vary by 
$0.88/f_U(8.6\,{\rm GHz})$~rad across 64~MHz
bandwidth at 1.4~GHz and $0.08/f_U(8.6\,{\rm GHz})$~rad at 2.5~GHz.   We 
searched for frequency-dependent variations in $V$
at all four frequencies by selecting two adjacent 32~MHz sub-bands at each 
frequency.  None were found.   Although the Faraday rotation 
(RM $\approx$ 69 rad/m$^2$)
across the band was clearly detected at 1.4 and 2.5~GHz, the variations in $V$ 
between sub-bands were less than 4\% and 0.8\% at these frequencies
respectively.  This result appears inconsistent with the derived lower 
limit on RRM, although the null result at 
1.4~GHz may result from an absence
of CP in the scintillating component (which in turn implies
$|U_0| /I \sim |V|/I < 0.01$ at 1.4~GHz).

The fact that all detections of the CP are of same sign also argues 
against this model.  If (i) $V$ does not change sign at any frequencies intermediate to those of our 
measurements (i.e. the spectrum is well-sampled) and (ii) $U_0$ does not 
change sign in the range 1.4$-$8.6~GHz then equation (3) implies 
$\lambda^3 {\rm RRM} < 2 \pi$ for 
all frequencies above 1.4~GHz (even if $V(1.4~{\rm GHz})$, whose sign 
is uncertain, is positive).
At 1.4~GHz one then has 
$|{\rm RRM}| < 6.2 \times 10^2 \, (1+z)^{3}$~rad/m$^3$, requiring 
$f_U(8.6\,{\rm GHz}) \geq 100$\% to be 
consistent with the lower limit on ${\rm RRM}$ obtained above.  While 
difficult to exclude entirely, we therefore conclude that the production of 
CP by a pair-dominated plasma is implausible.

Polarization conversion may also occur in a medium containing a 
mixture of
cold and relativistic pair plasma, in which case the fractional CP 
varies as $m_c \propto \nu^{-1}$ Pacholczyk (1973).  
This model appears implausible in light of the observed 
$\nu^{0.7_{-0.3}^{+1.4}}$ frequency 
dependence of the CP from 1.4 to 4.8~GHz.  

The presence of several distinct sub-components may 
alter the spectral properties of the observed CP.
However, the scintillation characteristics argue against the existence of 
multiple circularly polarized components, each with distinct $U_0$, RRM and 
$\gamma$.  The presence of multiple components with different $V/I$ 
would lead to substructure in the lightcurve of $V$ compared to the lightcurve
of $I$ as the scintillation selectively amplifies and deamplifies
parts of the source differently.  This would result in a loss of correlation 
between $V$ and $I$, particularly at 4.8 and 8.6~GHz, where the 
scintillation is most sensitive to small-scale structure.  This 
is not observed in fig. 1 where 
the correlation coefficients are close to unity.   However, it is more 
difficult to ascertain the presence of substructure in the variability 
at 1.4 and 2.5~GHz due to the long timescale of the fluctuations.
  
Jones \& O'Dell (1977a, 1977b)\markcite{Jones77a,Jones77b} 
presented a model for the CP of inhomogeneous 
synchrotron sources, incorporating optical depth effects, mode 
coupling and mode conversion due to the birefringence of the 
plasma.  Below the self-absorption turnover 
frequency, $\nu_{\rm SSA}$, mode coupling dominates, and the 
CP is typically less than 0.05\%, and certainly 
not more than 2\%.  Mode conversion dominates above $\nu_{\rm SSA}$, with the 
CP as high as 10\% near $\nu_{\rm SSA}$, and 
decreasing to less than $10^{-3}$ at frequencies a decade above
$\nu_{\rm SSA}$.  This model is viable only if $\nu_{\rm SSA}$ is
within a factor $\sim 1.4$ of the frequency at which the high (3.8\%) 
CP was observed, at 4.8~GHz.   This is difficult to 
verify as we do not know the intrinsic spectrum of the 
scintillating component and the frequency range of our observations is limited.

\placefigure{fig:FIG3N.PS}
\begin{figure*}
\begin{tabular}{c}
    \psfig{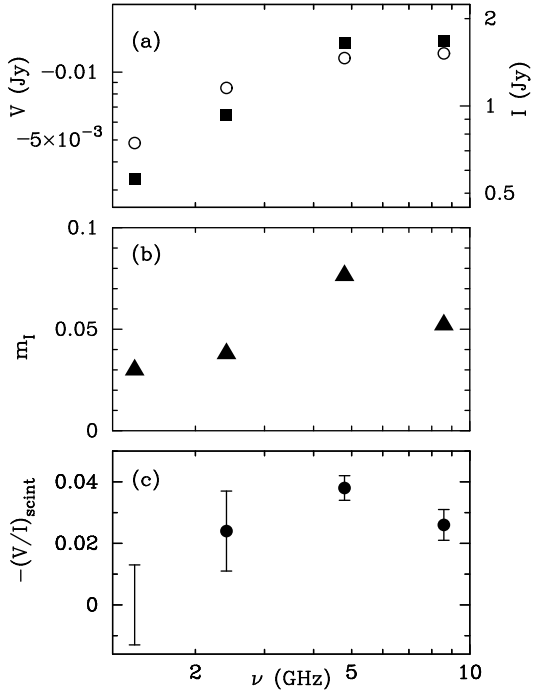}
\end{tabular}
\caption{Various quantities derived from the data in 
fig. 1.  (a) Spectra of the 
total intensity (circles) and CP (squares) in PKS~1519-273 for the four 
bands of ATCA averaged over the duration of the observations.
(b) Modulation indices derived from the intensity 
fluctuations over the 5 days of observations.  The long timescale of 
variability makes it difficult to compare these values with the 
ensemble-average quantities predicted by scintillation theory.  This 
is particularly relevant at 2.5 and 1.4~GHz because the timescale of 
variability is comparable to or exceeds the period of observation.
(c) The spectrum of circularly polarized component derived by 
comparing the magnitude of the fluctuations in $I$ with those in $V$ 
(see fig. 2).}
\end{figure*}

\section{Conclusion}
The variability detected in PKS~1519-273 in all four Stokes 
parameters at frequencies from 1.4 to 8.6~GHz is remarkable, but has a natural 
interpretation in terms of ISS.  The 
scintillation properties at 4.8~GHz constrain the brightness 
temperature of the scintillating component to $T_b \geq 5 \times 
10^{13}$~K, although there is strong evidence to suggest it may be as 
high as $6 \times 10^{14}$~K.  Comparison of the fluctuations in 
$I$ and $V$ imply that 
this component is exceptionally highly circularly polarized at 8.6 and 
4.8~GHz.  Simple applications of synchrotron theory and models of 
circular repolarization encounter difficulties with the spectral 
behavior and magnitude of the CP.  
The strong correlation between the fluctuations in $I$ and $V$ at 4.8 
and 8.6~GHz and the high sensitivity of the scintillation to source 
structure at these frequencies argue against a complex source, with different $V/I$ in each component.
Inclusion of effects due to small-scale inhomogeneity, mode coupling 
and optical depth effects may reproduce 
the observed characteristics of the CP.  However, 
this model is only viable if the frequency at which the
source is observed to become optically thin is in the range 
$3.4\,{\rm GHz} \lesssim \nu \lesssim 6.7\,{\rm GHz}$.  This possibility is 
presently difficult to confirm.  Even if correct, the puzzle 
remains as to why so few sources exhibit such high 
levels of CP.  

Finally, in light of the extremely high brightness temperature of 
PKS~1519-273, we advance the possibility that the observed emission 
is not due to synchrotron emission at all and that high CP 
may be a characteristic of a new emission mechanism.

\acknowledgments
We thank Don Melrose, Ron Ekers, Mark Walker, Jim Lovell, Dick 
Hunstead and Lawrence Cram for valuable discussions.  The 
Australia Telescope is funded by the Commonwealth Government for 
operation as a national facility by the CSIRO.

\end{document}